\documentstyle[psfig]{article}

\begin{document}


\title{The Onset of Mackey-Glass Leukemia at the Edge of Chaos}

\author{{M.~Argollo de Menezes and R.M.~Zorzenon dos Santos}\\
{Instituto de F\'{\i}sica, Universidade Federal Fluminense,}\\
 {Av.  Litor\^anea s/n,  24210-340,  Niter\'oi,  RJ,  Brasil} \\
 {e-mail:marcio@if.uff.br; zorzenon@if.uff.br}}


\maketitle

\begin{abstract}

In this paper we revisit the Mackey-Glass model for blood-forming process,
which was proposed to describe the spontaneous fluctuations of the blood
cell counts in normal individuals and the first stage of chronic
myelocytic (or granylocytic) leukemia (CML). We obtain the bifurcation
diagram as a function of the time delay parameter and show that the onset
of leukemia is related to instabilities associated to the presence of
periodic windows in the midst of a chaotic regime. We also introduce a
very simple modification in the death rate parameter in order to simulate
the accumulation of cells and the progressive increase of the minima
counts experimentally observed in the final stage of the disease in CML
patients. The bifurcation diagram as a function of the death rate
parameter is also obtained and we discuss the effects of treatments like
leukapheresis.

\end{abstract} 


\section{Introduction}

Many years ago Mackey and Glass \cite{Mackey_1977} proposed two different
simple models to describe chronic and acute diseases, generated by the
failure of physiological control systems, which may be identified by the
altered periodicity of some observable.  Recently Landa and Rosemblum
~\cite{Landa_1995} revisited one of the models proposed to describe the
respiration control and introduced a modification in the original equation
in order to take into account the activity of the brain on respiration. In
this paper we revisit the model proposed to describe the fluctuations in
the blood cell counts in normal individuals and the initial phase of
chronic myelocytic leukemia (CML).

 The human blood contains a remarkable variety of cells, each type
precisely designed to its own vital function.  All different types of
blood cells are produced in the bone marrow and all of them are originated
from earlier non-functional cells called {\it stem cells}.  The stem cells
replicate repeatedly and differentiate into various kind of secondary
cells.  The growth and development of cells in the bone marrow are
carefully controlled to produce the correct number of each type of cell
necessary to keep the body healthy.  The blood-forming process
(hematopoiesis) is subjected to active control by mechanisms acting to
buffer the system against disturbances and to provide adaptive
adjustment to non-stationary conditions ~\cite{The_cell}.  The
hematopoiesis is regulated by feedback control systems and as other
time-delayed processes ~\cite{Cherniak_1973,Glass_1979} it exhibits
oscillatory and cyclic behaviors.   

Most of the models proposed to describe physiological control systems
consider the differentiation steps of the secondary cells as different
compartments described by differential equations.  In general those models
have a high-dimensional parameter space (\cite{Rubinow_1975}-
\cite{Wheldon_1974}) which makes it quite difficult to compare theoretical
results to the experimental ones. In contrast to those more realistic, but
rather complicated models, Mackey and Glass ~\cite{Mackey_1977} proposed a
very simple model based on a single time-delayed differential equation,
with parameters estimated from experimental data. The change of the time
delay parameter leads to different kind of oscillations of the cell
counts which may be associated to healthy and non-healthy behaviors when
compared to experimental data obtained by Gatti et al \cite{Gatti_1973}.  
The onset of an abnormal dynamics, which may be associated to the onset of
leukemia is then obtained in this model by the gradual tuning of a
control (time-delay) parameter. Since its proposal this model has been
mostly used as a paradigm for non-linear systems (\cite{Kazarinoff_1979}-
\cite{Berre_1987}),and less attention was paid to its biological
applications. 

In this paper, we show that this simple model that generates very complex
dynamical behaviors for cell counts, is more robust than thought to
reproduce the onset of leukemia and it is also possible to use it to
reproduce the two phases of leukemia and the effects of some treatments.
The observed instabilities that occur at the onset of leukemia can be
obtained for different values of the time delay parameter, corresponding
to different possibilities of the system to achieve the edge of chaos. In
other words, the instabilities resulting from complex interactions
involved in the physiological control systems that may describe the
oscillations observed in CML patients correspond to an state hovering
between the incoherence of chaos and the spontaneous order found in
periodic behaviors (\cite{kauffman}).

In section 2 we present the biological motivation that inspired the model
discussed here and other experimental data concerning the discussions we
present along the paper. In section 3 we introduce the model. In section 4
we present the results obtained. We first obtain the bifurcation diagram
in function of the time delay parameter for the Mackey-Glass model and
show that the onset of leukemia is related to instabilities associated to
transition regions between chaotic and periodic regimes, usually refereed
in the literature as the edge of chaos.  These transition regions occur
for a set of different values of the time-delay parameter and not only for
a unique value of this parameter. The bifurcation sequence observed in
this diagram is quite uncommon and presents only some of the periods
suggested by the authors \cite{Mackey_1977,Glass_1979}. We also introduce
a simple modification in the death rate parameter of this model in order
to simulate the progressive increase of the minima counts in phase II of
CML. The bifurcation diagram in function of the death rate parameter is
also obtained and the effects of leukapheresis treatments is also
discussed. Our conclusions are summarized in section 5.

\section{The Biological Motivation}

It has been observed spontaneous fluctuations with periodic behavior in
the peripheral blood cell counts of healthy individuals, in chronic
myelocytic leukemia (CML) patients \cite {Morley_1967,Vodopick_1972} as
well as in other pathological states \cite{Glass_1979}. In healthy
individuals constant or mild oscillations of the peripheral-blood
leukocyte-counts occur with periods varying from 14 to 24 days
~\cite{Morley_1967}, while in many patients with CML these oscillations
have periods varying from 30 to 120 days
(\cite{Morley_1967,Kennedy_1970}), with a cell accumulation process
observed in the final stage (phase II) of the disease. Morley et al
\cite{Morley_1967} suggested that in patients with cyclic leukocytosis and
CML, the bone marrow is either unduly sensitive to feedback stimulus or
being excited intermittently by an abnormal stimulus, generated outside
the marrow.

 From long term observations Gatti et al. ~\cite{Gatti_1973} showed that
the onset of leukemia (CML) is characterized by two distinct phases: phase
I, a stable 15-month period, from time of diagnosis, during which
leukocyte counts peaked every 72 days and then returned to the baselines
level; and phase II, a 18-month period during which the minima levels of
leukocyte counts became progressively higher indicating that those cells
were accumulating in the peripheral blood.  According to them, the
increased marrow production in the disease does not represent a
life-threatening situation, until cell accumulation begins. During phase
I, the leukocyte counts return to baseline levels, indicating that the
overproduction of cells in the peak of a given cycle could be cleared from
the peripheral blood before the beginning of the next cycle.  In contrast,
in phase II, the leukocytes accumulate in the peripheral blood of the
patient, suggesting primarily an increase in the cell production or a
progressive loss of cell-clearing capacity with little or no changes in
the production capacity \cite{Gatti_1973}.

Patients with CML were submitted to leukapheresis treatments in order to
test out if this kind of treatment would be able to reverse the rising
leukocyte levels providing a good therapy to maintain the patient healthy
for years. The leukapheresis consists of 8 weeks of intense treatment to
remove about $30-50\%$ of the mature and premature leukocytes from marrow
and peripheral blood. According to the results obtained \cite{Gatti_1973},
those treatments are able to retrieve the system from the behavior of
phase II to the behavior observed in phase I, with low amplitude cycles
and minima counts. However, after one year, subsequent cycles (of 72 days
each) indicated re-accumulation of cells in the peripheral blood.  In
other words, these studies clearly demonstrated that leukapheresis is
effective for the rapid reduction of elevated cell counts, but long-term
treatment has not improved survival of the patients with CML, since this
kind of treatment do not alter the mechanisms underlying the disease
process.

Nowadays this kind of treatment is not used anymore, and have been
replaced by other treatments like chemotherapy and radiotherapy that in
some cases may lead to the patient cure. The mechanisms underlying the
effectiveness (or not) of those modern treatments are not clear-they act
by producing an intense immunosupression followed by an intense activation
of the bone marrow in order to recover the system, involving different
kinds of drugs- and it would be impossible to discuss its effects on this
very simple model. However, it is possible to analyze and simulate the
effects of leukapheresis in the MG model as will be shown in section 4.

\section{The Model}

The Mackey-Glass model \cite{Mackey_1977} relates the number of precursor
and mature cells resident on the bone marrow to the number of cells
released into the blood stream.  The time evolution of an homogeneous
population of mature, circulating cells of density $P$ is then given by:

\begin{equation} \label{eq:MG} \frac{dP(t)}{dt} ~=~ \frac{{\beta}_0
{\theta}^n P_{\tau}}{{\theta}^n + P_{\tau}^n } - \gamma P(t) 
\end{equation}

where the first term in the right hand side corresponds to the cellular
production, which is described by a (Hill) function of
$P_{\tau}=P(t-\tau)$, where $\tau$ represents the delay between the production
of the precursor cells and the release of the mature cells into the blood.
The average (density independent) death rate of the population is given by
$\gamma$.  The constant values of $\beta_0$, $\theta$, $n$ and $\gamma$ are
estimated from experimental data \cite{Mackey_1977}.

Like other theoretical models (\cite{Rubinow_1975}-\cite{Wheldon_1974}),
the Mackey-Glass model associates the onset of cyclic leukocytosis and
leukemia with instabilities in the dynamics of the model.  They found that
for $\tau=6$ days the solution has low amplitude oscillations with period
of 20 days, which describes qualitatively the mild oscillations observed
in the circulating levels of leukocytes in normal healthy adults
\cite{Morley_1967}.  For $\tau = 20$, the numerical solution reproduces
the altered periodicity observed in phase I of the diagram of circulating
leukocytes as a function of time of patients with diagnosed CML
~\cite{Gatti_1973}.

\section{ Results}

\begin{figure} 
 \centerline{\psfig{file=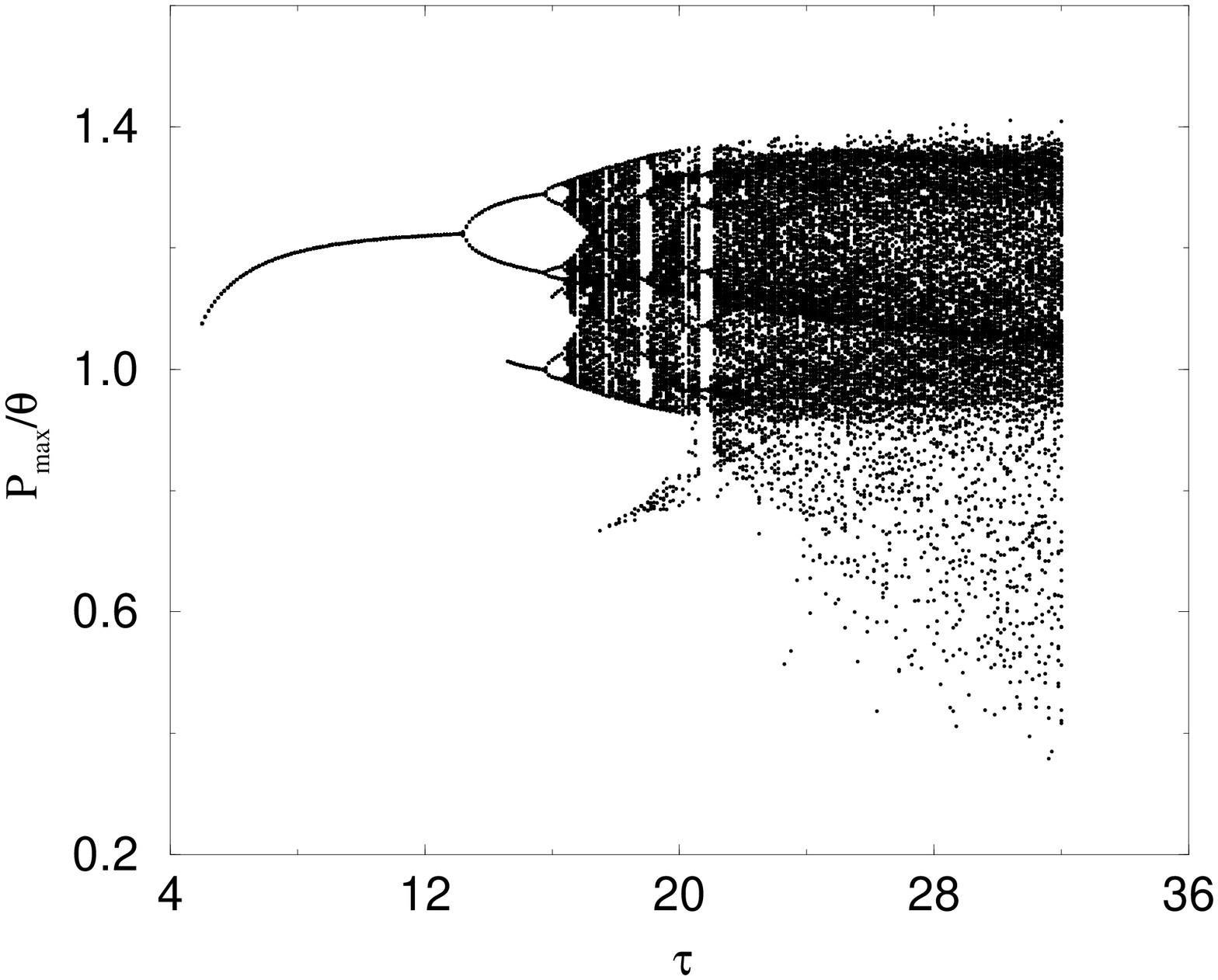,height=10cm,angle=270}}
 \caption{Bifurcation diagram as a function of the time delay parameter
 $\tau$, for the Mackey-Glass time delayed differential equation using
 $\beta_0=0.2$ per day, $\theta= 1.6 \times 10^{-6}$, $n=10$ and
 $\gamma=0.1$.}  
\label{menezes1}
\end{figure}

By standard fourth-order Runge-Kutta integration techniques the numerical
solution for the equation is obtained, where we adopt the same set of
parameters of the original model ~\cite{Mackey_1977}.  In order to
characterize the dynamical instabilities associated to the onset of
leukemia we have obtained the bifurcation diagram of the local maxima of
the time evolution of $P(t)$ as a function of the time delay parameter
$\tau$ as shown in figure \ref{menezes1}. This diagram was built from the
analysis of the cross section of the return map for different values of
$\tau$. By increasing the time-delay parameter the oscillations become
unstable and new cycles with different periods appear, leading to
aperiodic and then to chaotic behavior. In contrast with the findings of
Mackey-Glass, we found a quite uncommon bifurcation sequence: $1$,
$2$, $3$, $6$, $7$, $14$, etc. with windows of period $6$ and $9$ in the
midst of the aperiodic regime.  We also note the appearance of a branch of
sparse points just above $\tau=17$.  This branch persists when we vary the
initial conditions.

We have observed that the very same altered periodicity observed for
$\tau=20$ (which was  associated to the onset of the cyclic leukocytosis
by Mackey and Glass \cite{Mackey_1977}) could also
 be reproduced for other values of $\tau$, for instance, $\tau= 18.6$ and
$21.2$. The common feature among these values of the time delay
parameter is that they correspond to the edge of the periodic windows
which appear in the midst of aperiodic regime.  This indicates the
robustness of the model to describe the disturbances that would correspond
to the onset of leukemia.

\begin{figure} 
 \centerline{\psfig{file=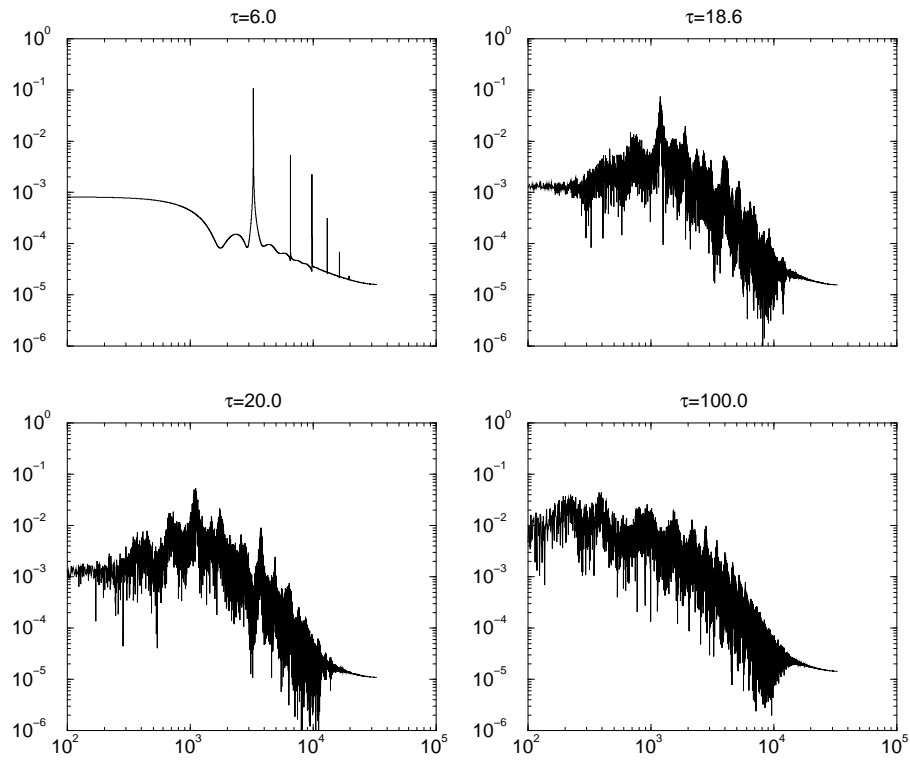,height=10cm,angle=270}}
 \caption{Power spectra of the numerical solutions of the MG equation for
 periodic ($\tau=6$), quasi-periodic ($\tau=18.6$ and $\tau=20$) and chaotic
 ($\tau=100$) regimes.}
\label{menezes2}
\end{figure}

In this model, the regular behavior of the blood-forming process and the
pathological states of disease correspond to different attractors of the
dynamics.  The standard methods of characterization of the attractors can
be applied for small values of $\tau$ (periodic behavior), whose
attractors have low embedding dimension, and for large values of this
parameter (chaotic behavior), whose embedding dimension of the attractors
can be approximated to infinity, for which simple asymptotic solutions
exist ~\cite{Kazarinoff_1979,Mensour_1995}.  Nevertheless, for
intermediate values of this parameter the attractors are embedded in a
high-dimensional space and the characterization becomes quite difficult,
specially in cases like this one, which involves time-delayed information.  
As far as we know up to now the available techniques do not allow a
precise characterization of the dynamics for intermediate values of
$\tau$. In figure \ref{menezes2} we show the power spectra for four values
of
$\tau$ corresponding to the three different dynamical behaviors described
above. Our results indicate that the special values corresponding to the
onset of leukemia would correspond to a quasi-periodic behavior: the power
spectrum has numerous peaks and presents a slow decay to zero. This
behavior should correspond to the zero value of the Lyapunov exponent.
Nevertheless, because of the difficulties mentioned above, the
calculations performed are still not conclusive.

\begin{figure} 
\centerline{\psfig{file=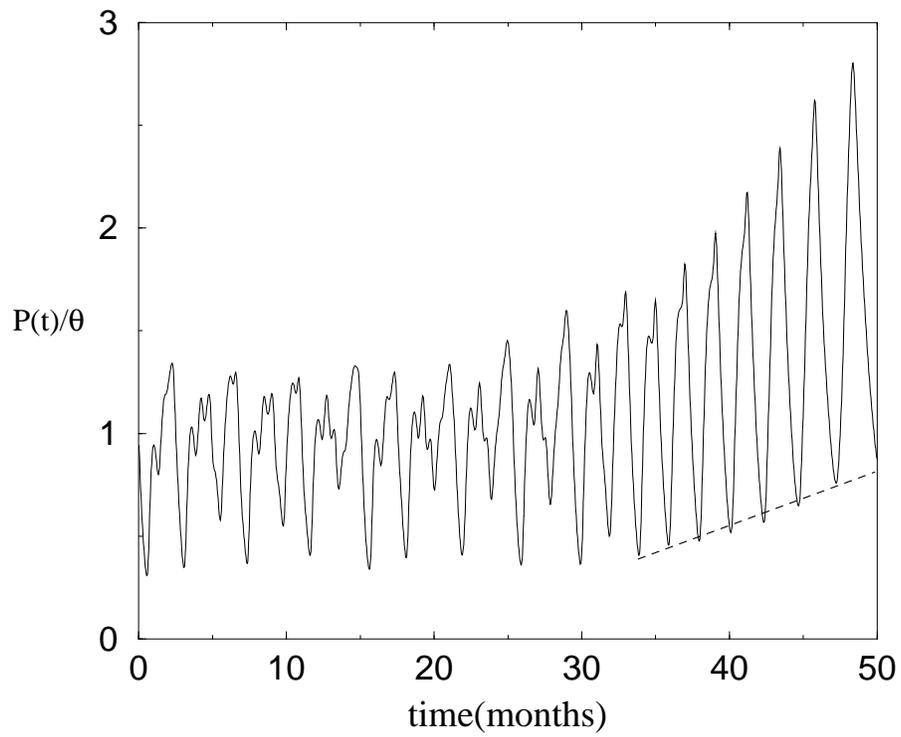,height=10cm,angle=270}}
\caption{Numerical solution showing qualitative behavior very similar to
phases I and II of leukemia \protect\cite{Gatti_1973}.}  
\label{menezes3}
\end{figure}

As mentioned before, the phase II of the common pattern observed in CML
patients may be associated to either a failure of the cell-clearing
capacity of the system or a over-production of immature cells (the cells
lose the ability to mature and differentiate) which leads to the
accumulation of cells and to a progressive increase of the minima counts
in the leukocyte population ~\cite{Gatti_1973}. In order to simulate the
phase II, we introduce a simple modification to the death rate that would
correspond to a failure in the cell-clearing capacity. Modifications in
the cell production, in order to simulate the loss of the cell's ability
to mature and differentiate is not straightforward and will be discussed
elsewhere. The different dynamical behaviors observed in the two distinct
phases, will be related to changes in the death rate, since in phase II it
will be no longer constant. The death rate $\gamma$ will be described
during phase II by a step function that diminishes by $10^{-4}$ every $60$
days. This simple modification simulates a system that slowly loses the
ability to remove cells, and reproduces qualitatively the increase of the
minima counts, observed by Gatti et al ~\cite{Gatti_1973} in phase II of
the disease, as shown in figure \ref{menezes3}.  We have also tried other
possible functions, but this one was more appropriate to describe the
accumulation of cells observed in phase II.

Figure \ref{menezes3} is obtained for the same set of parameters used in
the original model. We have arbitrarily chosen to let the system evolve
with $\gamma$ constant until $t=4900$ (which would correspond to the end
of phase I) and after this point the death rate begins to decrease. The
slope of the linear growth of the minima counts in our simulations has the
same order of magnitude as the equivalent slope in the experimental data.

\begin{figure}
\centerline{\psfig{file=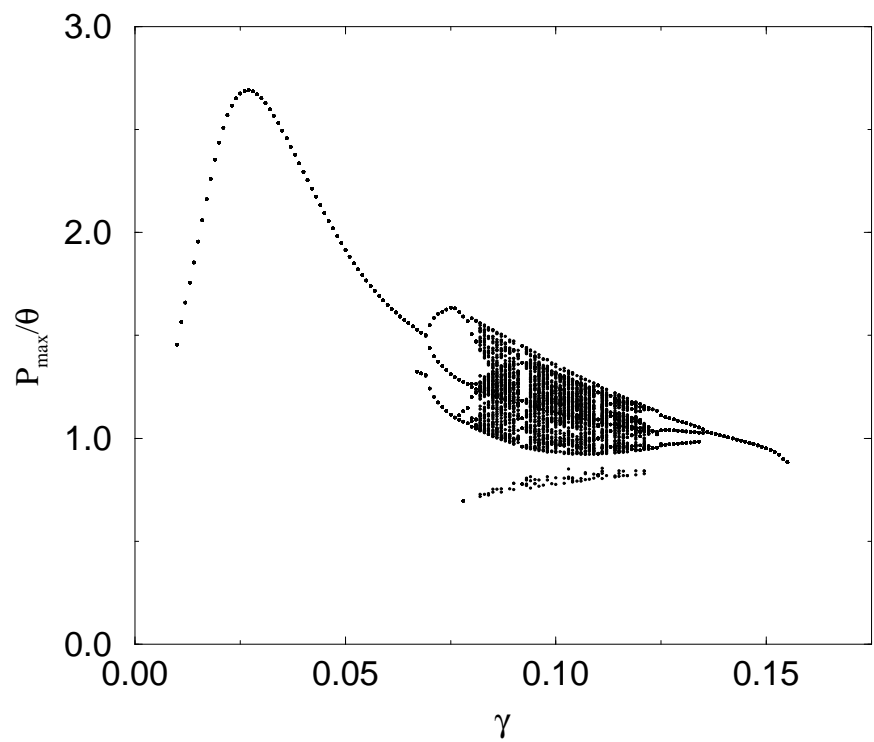,height=10cm,angle=270}}
\caption{Bifurcation diagram as a function of $\gamma$, obtained for the
same set of parameters of figure 1, for $\tau=20$.} \label{menezes4}
\end{figure}

In figure \ref{menezes4} we present the bifurcation diagram as a function
of $\gamma$, for $\tau = 20$, in the range of interest. The diagram shows
a bifurcation sequence which ends up in a limited region from $0.08$ to
$0.12$ of aperiodic behavior, above which the system becomes periodic
again.  In the midst of the aperiodic region we also find windows of
periodicity $6$.

\begin{figure} 
\centerline{\psfig{file=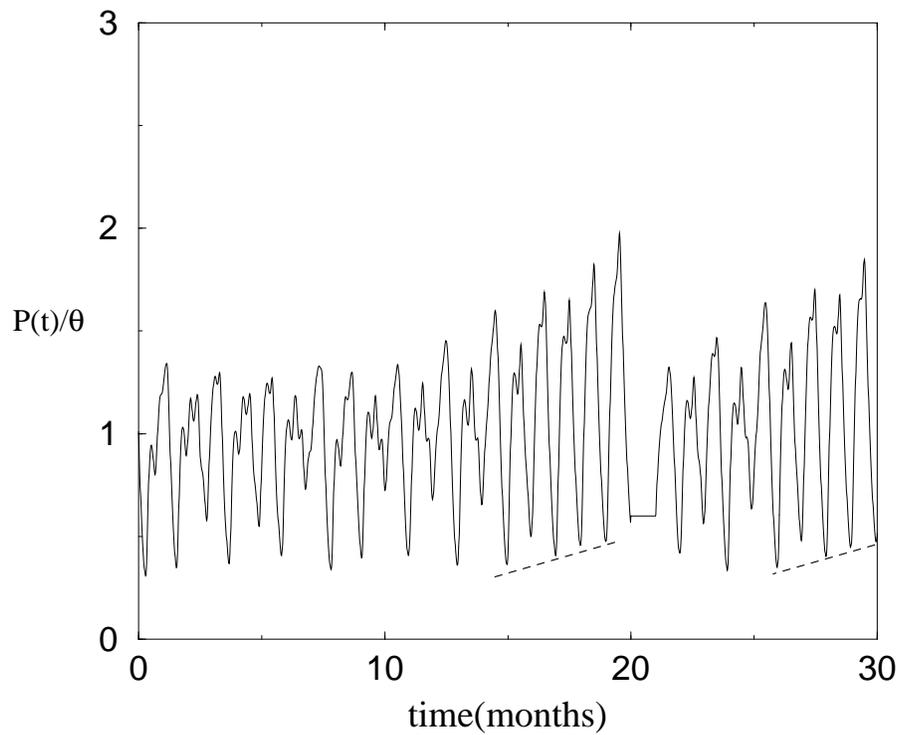,height=10cm,angle=270}}
\caption{Simulation
of two Leukapheresis (I and II)  considering the same set of parameters of
figure 1, with a time interval of $400$ days between the two treatments.
Since the disturbance continues to act after the
treatment, after some time we observe the same increase of the minima
densities, as observed in the experimental data shown in figure $1$
of ref. $9$.} 
\label{menezes5}
\end{figure}

We also simulate the effects of leukapheresis treatments during the phase
II. We let the system evolve during some time ($600$ days) in phase II and
then we simulate the first leukapheresis, in order to test if this
cleaning procedure would revert the raising of the minima counts. The
treatment is simulated maintaining the density of cells constant during 60
days (equivalent to the 8 weeks of real treatment) in a level that
corresponds to $60\%$ of the value of the density in the beginning of the
protocol~\cite{Morse_1966}.  For $\tau = 20$, after 60 days the system
does not have memory of what happened before the treatment.  After each
treatment we let the system evolve under the same conditions, since the
treatment only removes the excess of cells accumulated and do not
interferes on the underlying mechanisms that could generate the
disturbance. Just after the treatment the minima counts returned to
the
baseline levels, but after a while we have observed the same linear
increase of the minimal counts of phase II, as in the experimental data.
The numerical solutions obtained for two leukapheresis treatments with
approximately one year interval ($400$ days) between each other is shown
in figure \ref{menezes5}. The results show a good qualitative agreement
with the experimental results \cite{Gatti_1973}. Before the first
treatment and between the two treatments $\gamma$ is practically constant
(varies approximately from $0.1$ to $0.099$), maintaining itself in the
aperiodic region of the $\gamma$-bifurcation diagram , with no change of
attractor (see figure \ref{menezes4}).

\section{Conclusions} 

The Mackey-Glass model for the blood-forming process reproduces the main
features observed in the peripheral blood-cell counts for normal
individuals and pathologies like cyclic leukocytosis and the disorder that
characterizes the accumulation of cells observed in a patient diagnosed
with leukemia (CML) in phase I and II. We have shown that the onset of
leukemia is related to special intermediate values of the time-delay
parameter, which seem to correspond to quasi-periodic behaviors. From the
bifurcation diagram as a function of the time-delay parameter $\tau$, we
observed that the onset of leukemia is associated to the edge of periodic
windows in the midst of aperiodic behavior. By varying the death rate
parameter we could reproduce the phase II behavior observed in the
experimental data, and with a simple cell-removing procedure we simulated
the effects of leukapheresis.  We have also obtained the bifurcation
diagram as a function of the death rate $\gamma$.

\section{Acknowledgements}

We thank A.P. Serbeto, J. Stilck and A. Bernardes for the critical reading of
this manuscript. This work was partially supported by the Brazilian Agencies
Conselho Nacional para o Desenvolvimento Cient\'{\i}fico e Tecnol\'ogico -
CNPq, CAPES and FINEP.


\newpage 

\end{document}